\documentclass[apj]{emulateapj}
\setkeys{Gin}{width=0.48\textwidth}
\usepackage{graphicx}
\newcommand{\hmpc}{\ifmmode{h^{-1}\,\hbox{Mpc}}\else{$h^{-1}$\thinspace Mpc}\fi}

\newcommand{\kms}{\ifmmode{\,\hbox{km\,s}^{-1}}\else {\rm\,km\,s$^{-1}$}\fi}
\newcommand{\msun}{{\rm\,M_\odot}}

\begin{document}
\title{The Dynamics of Star Stream Gaps} 
\shorttitle{Dynamics of Star Stream Gaps}
\shortauthors{Carlberg}
\author{R. G. Carlberg}
\affil{Department of Astronomy and Astrophysics, University of Toronto, Toronto, ON M5S~3H4, Canada} \email{carlberg@astro.utoronto.ca }

\begin{abstract}
When a massive object crosses a star stream velocity changes are induced both along and
transverse to the stream which can lead to the development of a visible gap.
For a stream narrow relative to its orbital radius the time of stream crossing is sufficiently short that the impact approximation 
can be used to derive the changes in angular momenta and radial actions along the star stream. 
The epicyclic approximation is used to calculate the evolution of the density of the stream as it orbits around in a galactic potential.
Analytic expressions are available for a point mass, however, the general expressions are easily
numerically evaluated for perturbing objects with arbitrary density profiles.  
With a simple allowance for the velocity dispersion of the stream, moderately warm streams can be modeled. 
The predicted evolution agrees well with the outcome of simulations of stellar streams for streams with widths 
up to 1\% of the orbital radius of the stream.  
The angular momentum distribution within the stream shears out gaps with time, 
further reducing their visibility, although the size of the shear effect requires more detailed simulations. 
An illustrative model indicates that shear will  limit the persistent gaps to a minimum length of a few times the stream width.
In general the equations are useful for dynamical insight into the development of stream gaps and their measurement.
\end{abstract}

\section{INTRODUCTION}
\nobreak

Star streams are created as the gravitational field of the galaxy tidally disrupts a dwarf galaxy or globular star cluster
orbiting in the halo of a galaxy.
Streams will exhibit a rich range of features along their length as a result of variations in the gravitational field.
A progenitor in any non-circular orbit will experience varying tidal fields around the orbit, leading to an
increased mass loss rate near perigalacticon \citep{DOGR:04}. 
As the unbound stream stars orbit they bunch up at apogalacticon and spread out at  perigalacticon \citep{Johnston:98, JSB:01}.
Stars leave the progenitor near the inner and outer Lagrange points where
unbound stars are roughly collimated into a radial outward stream which leads to regular cycloidal density variations 
near the beginning of the stream  \citep{Kupper:08,Kupper:10,Kupper:12}. 
Finally, within an LCDM dark matter halo there are predicted to be large numbers of dark matter sub-halos \citep{VL1,Aquarius,Stadel:09} with $n(M)\,dM \propto M^{-1.9}\,dM$.
The sub-halos induce gaps when they cross a star stream \citep{YJH:11,Carlberg:09,Carlberg:12}.  

Although gaps are purely a statistical measure of the dark matter sub-halo population they do have the attractive property that the
individual cross-section depends on their radius, $R\propto M^{0.43}$, 
which means that the cross-section for gap creation integrated over the sub-halo population, 
$A(M)\, n(M)\, dM$, is dominated by the smallest masses that can create a visible gap. 
The dominance of the low mass members of the population is useful in discriminating between cold and some forms of warm dark matter, where warm dark matter has very few sub-halos below $\sim10^{8-9}\msun$ for a 1 Kev dark matter particle \citep{BHO:01,Bode:01,Benson:13,AHA:13,SSR:13}.

In this paper we develop the general expressions for the changes in velocities as a result of the passage of massive object. 
We re-derive the result of \citet{YJH:11} for the velocity change along the stream and present the 
the result perpendicular to the stream. 
We use the general  expressions for an arbitrary mass density profile
 to derive the changes for a stream on a galactic orbit  and the resulting evolution of density along the stream.
Adding a dispersion in angular momentum and radial action allows for the finite width of streams.
The predictions are then compared to velocity changes and gaps in simulations.

\section{Response of the Stream Stars to an Encounter}

We make a few assumptions to simplify the analysis of the encounter between a massive object and a stream of stars. 
First, we assume that the encounter is sufficiently fast and weak that the changes in the star stream velocities are small relative to their orbital velocities,  {\it i.e.}, the impact approximation. 
Second, we will do the basic analysis for a zero initial width stream. However, we will show that the results 
are applicable to streams of finite width with random velocities and shear velocities. 
And third, we assume that the orbit of the stream is circular in an axisymmetric potential.
In a non-circular orbit there will be an additional systematic compression and extension of the stream and the features within it around the orbit.

\subsection{Velocity Changes of the Stars}

\begin{figure}
\begin{center}
\includegraphics[angle=0, scale=0.7]{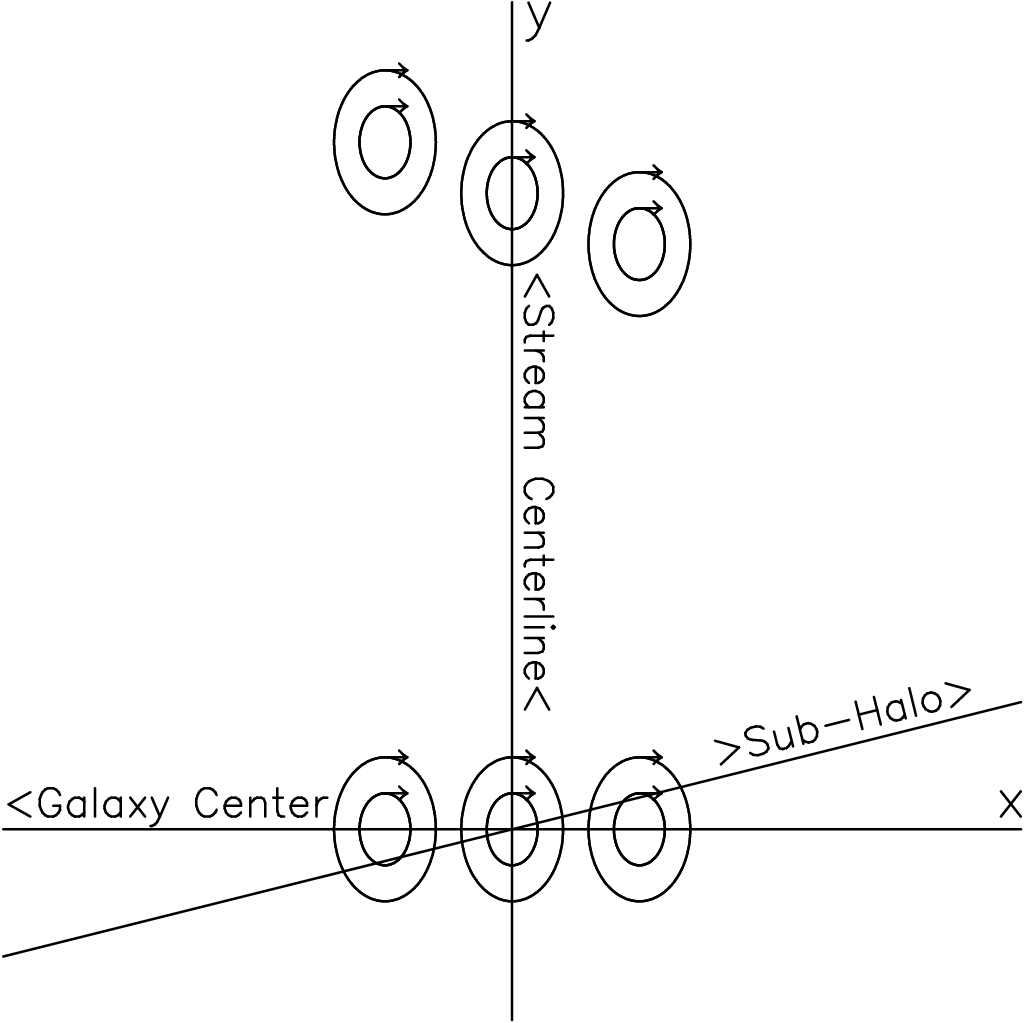}
\end{center}
\caption{The coordinate system for the analysis and the basic motions. The stream moves upward at velocity $v_y$ along the $y$  axis with 
the $t=0$ value of $y$ being used to label the stars. The sub-halo crosses the x-axis at $[b_x,0,b_z]$ at $t=0$ 
with a velocity $[V_x, V_y, V_z]$.
The epicycles rotate opposite to the stream rotation around the galaxy. 
The angular velocity decreases outwards so stars with  guiding centers at larger radii fall behind those at smaller radii. }
\label{fig_coords}
\end{figure}

The epicyclic approximation expands about an orbit of constant angular momentum, 
$J_\phi$, where $\phi$ is the angle around a circular orbit.
The local co-ordinate system for our analysis is shown in Figure~\ref{fig_coords}.
A Cartesian frame with the $x$ coordinate is aligned in the radial direction \citep{BT:08} measured
from the center of the progenitor. 
The stream then moves in the positive $y$ direction. 
Stars with the same angular momentum share a common guiding center radius, but differences in radial velocity lead to different size epicycles around the guiding center. 
Stars with larger angular momentum have a larger guiding center radius where the rate of 
angular rotation is lower, so stars initially aligned 
at the same $y$ but at different $x$ will gradually shear with respect to each other as they move forward along the $y$ axis at different 
rates.

The section of the stream that interacts most strongly with a passing sub-halo is a few times the sub-halo's scale radius
\citep{Carlberg:12}.
For sub-halos of $10^8 \msun$ the scale radius is about 0.8 kpc so the interaction region for a sub-halo orbiting in the halo beyond the disk  is normally
sufficiently small relative to the orbital radius that the response to the perturbation
can be calculated as if the stream were a straight line.
The coordinate distances between the stream stars moving at a uniform velocity $v_y$ and a sub-halo 
moving at velocity $[V_x, V_y, V_z]$ which crosses $y=0$ at time $t=0$ is,
\begin{equation}
\vec{d}(y, t) = [b_x+V_x t, y+ (v_y-V_y)t, b_z+V_z t],
\label{eq_d}
\end{equation}
where $y$ is the $t=0$ location of stars along the stream and  the closest approach 
distance $b$ is resolved into its  $x$ and $z$ components. 
A perturbing mass with a spherically symmetric gravitational potential  $\Phi(r)$,
 induces a net  velocity change along the stream of,
\begin{equation}
\Delta\vec{v}(y) = -\int_{-\infty}^{\infty}  \vec{\nabla}\Phi(| \vec{d}(y,t)|)\, dt.
\label{eq_dv}
\end{equation}
Equations~\ref{eq_dv} are straightforward to numerically integrate for most radially symmetric density profiles.

To illustrate the behavior of these equations the velocity changes of Equations~\ref{eq_dv} can be analytically integrated for a perturbing point mass, $M$. 
We define the velocity of the point mass relative to the stream, $v_\parallel=v_y-V_y$ and
orient the $x-z$ frame so that $v_\perp$ is the velocity toward the stream, $\sqrt{V_x^2+V_z^2}$.
The impact parameter is $b=\sqrt{b_x^2+b_z^2}$.   

The substitutions allow the integral of Equation~\ref{eq_dv} for the component of the velocity change
parallel to the stream motion for the point mass to be written as,
\begin{equation}
\Delta v_\parallel(y)= \int_{-\infty}^{\infty}
{\frac {{\it -GM}\, \left( v_\parallel t+y \right)}{ \left[ {v_\perp}^{2}{t}^{2}+ \left( v_\parallel t+y \right) ^{2
}+{b}^{2} \right] ^{3/2}}}\, dt.
\label{eq_idvpar}
\end{equation}
Doing the integral of Equation~\ref{eq_idvpar} gives the change in the $v_\parallel$ component of the stream stars is,
\begin{equation}
\Delta v_\parallel(y)={\frac {-2{\it GM}{v_\perp}^2 y}
{v \left( v^2 b^2+{v_\perp}^2 y^2 \right)}},
\label{eq_dvpar}
\end{equation}
where $v= \sqrt{v_\parallel^2+v_\perp^2}$ is the speed of the perturbing mass relative to the stream stars.  
This equation has been previously derived in \citet{YJH:11} with slightly different notation.

Perpendicular to the stream the velocity change is,
\begin{equation}
\Delta v_\perp(y)= \int_{-\infty}^{\infty}
{\frac {{\it -GM}\,v_\perp t }{ \left[ {v_\perp}^{2}{t}^{2}+
 \left( v_\parallel t+y \right) ^{2}+{b}^{2} \right] ^{3/2}}}\, dt,
\label{eq_idvperp}
\end{equation}
which integrates to,
\begin{equation}
\Delta v_\perp(y)={\frac {2{\it GM}\,v_\perp v_\parallel y}
{v \left( v^2 b^2+{v_\perp}^2 y^2 \right)}}.
\label{eq_dvperp}
\end{equation}
Equation~\ref{eq_dvperp} is non-zero along the stream only in the special case that $v_\parallel=0$.

It is interesting to note that the ratio of Equations~\ref{eq_dvpar} and \ref{eq_dvperp} is,
\begin{equation}
{{\Delta v_\perp(y)}\over {\Delta v_\parallel(y)}}=-{v_\parallel \over v_\perp}.
\label{eq_ratio}
\end{equation}
The ratio of the two velocity perturbations does not depend on distance along the stream or impact parameter, only on 
the velocity of the perturbing mass  relative to the stream. 
We note that since $v_\parallel$ is the relative velocity in the stream direction of the perturber and the stream stars 
it can be small and  even change sign. 

\subsection{Orbital Changes and Stream Gaps}

The stream is orbiting within a galaxy, hence the 
response of the stream to the perturbing mass needs to be placed within the framework of orbital dynamics. 
For simplicity we continue to assume that the stream
is on a circular orbit at a radius $X_0$. The relation between the linear 
coordinate $y$ and the $t=0$ angular coordinates, $\phi(0)$, is simply
$y=\phi(0) X_0$.
The velocity change parallel to the motion of the stream, $\Delta v_\parallel(y)$, 
are changes in the angular momentum, $J_\phi$, of the stream stars.  For small ellipticity orbits
the epicyclic approximation gives the angular momentum as $J_\phi=X v_c$, where  $X$ is
the guiding center radius and $v_c(X)$ is the circular velocity. The guiding center rotates forward at a uniform rate $\Omega(X)$.
The epicycle is centered on the guiding center, with the star traveling around the 
elliptical epicycle in the opposite direction to the mean angular rotation at a rate $\kappa(X)$ \citep{BT:08}.
After the encounter with a sub-halo the angular momentum along the stream is,
\begin{equation}
J_\phi(y) = v_c(X_0) X_0 + \Delta v_\parallel(y) X_0,
\label{eq_J}
\end{equation}
where $X_0$ is the pre-encounter guiding center radius.
The angular momentum changes lead to new guiding centers along the stream,
\begin{equation}
X=X_0 {{v_c(X_0)+\Delta v_\parallel(y)}\over v_c(X)}.
\label{eq_r}
\end{equation}
Consequently, after the encounter the stars move at angular rates that are a function of their new guiding centers,  $\Omega= v_c/X$,
or,
\begin{equation}
\Omega(y) = {v_c\over X_0}\left[1 + {\Delta v_\parallel(y) \over{v_c}}\right]^{-1},
\label{eq_Om}
\end{equation}
where we have assumed for simplicity that $v_c$ is locally constant, although the result can be generalized to any rotation curve shape using a linear expansion.
An acceleration forward leads to the stars rotating more slowly and vice-versa, which opens up a gap in the star stream centered 
on $y=0$, the crossing point of the sub-halo.

\subsubsection{The Gap Density Profile}

The length and approximate density profile of a gap are readily calculated from the perturbed motion of the stars.
The linear density along the stream is $\rho \equiv X_0^{-1}dn/d\phi$. 
After the encounter the stars in the stream are at angles $\phi (t)= \phi +\Omega(\phi X_0) t$. 
Therefore the differential in the density equation becomes, 
\begin{equation}
d \phi(t) = \left(1+X_0{ d \Omega(y)\over {d y}}\, t\right) d \phi.
\label{eq_dphi}
\end{equation}
Using Equation~\ref{eq_Om} in Equation~\ref{eq_dphi}
 the linear density along the stream becomes,
\begin{equation}
\rho(y,t) = \rho_0 \left[1- \left(1+{{\Delta v_\parallel(y)}\over v_c}\right)^{-2} {{d \Delta v_\parallel(y)} \over{dy}}\, t\right]^{-1}.
\label{eq_den}
\end{equation}
In principle one could allow for streams on non-circular orbits by including the variation of $\Omega(r)$. 

Equation~\ref{eq_den} gives the density as a function of time, labeled with the initial $y$ values. 
The gap has its greatest depth at $y=0$, where Equation~\ref{eq_dvpar} gives $\Delta v_\parallel=0$ 
its derivative is always negative for reasonable perturbing mass profiles.
Therefore the density in the gap goes to zero asymptotically as $t^{-1}$. 
On either side of the gap where $d\Delta v_\parallel(y)/dy>0$ Equation~\ref{eq_den} will
fail for sufficiently large $t$ values that the expression in brackets goes through zero and becomes negative.

As the guiding centers change in response to the angular momentum changes,  
the mean orbital angle at which the density applies is modified from the pre-encounter $\phi(t)=y/X_0+\Omega_0 t$, to,
\begin{equation}
\phi(y,t) = {y\over X_0} + \Omega(y) t,
\label{eq_phit}
\end{equation}
where we use the $\Omega(y)$ of Equation~\ref{eq_Om} to give the result. 
Equation~\ref{eq_phit} ignores the epicyclic oscillations of individual stars,
on the basis that this oscillating term will normally average to zero in a realistic stream that contains random motions.
Equation~\ref{eq_phit} is the angle relative to the current location of the middle of the gap.
Subtracting the gap center angular rotation of $v_c/X_0$ from the $\Omega(y)$ of Equation~\ref{eq_Om}
removes the mean motion to give $\Delta\phi(y,t)=\phi(y,t)-v_c/X_0$, which when into Equation~\ref{eq_phit} gives, 
\begin{equation}
\Delta \phi(y,t)
\label{eq_dphit} =  {y\over X_0} - {v_c\over X_0}{\Delta v_\parallel(y) \over{v_c}}\left[1 + {\Delta v_\parallel(y) \over{v_c}}\right]^{-1} t.
\end{equation}
Equations~\ref{eq_den} and \ref{eq_dphit}
are parametric equations in the density, $\rho(y,t)$ and angle $\Delta\phi(y,t)$ with $y$  and $t$ as parameters.

The density in the stream, Equation~\ref{eq_den}, develops a gap in the region where $d \Delta v_\parallel /d y$ is negative.
 For the point mass object,
\begin{equation}
{d \Delta v_\parallel \over {d y}} = - {{2v_\perp^2 {\it GM}(v^2b^2 - v_\perp^2 y^2)}
	\over{v (v^2b^2 + v_\perp^2 y^2)}}.
\label{eq_dvp}
\end{equation}
Therefore the stream will have a density reduction over  an interval of length given by 
the locations of the zeros of  
$d \Delta v_\parallel /{d y}$,
or  $y=\pm v/v_\perp b$ at time $t=0$ of Equation~\ref{eq_phit}.
Equation~\ref{eq_Om} gives the second term of Equation~\ref{eq_phit}.
The time development of the gap length in our point mass example then is,
\begin{equation}
\ell(t) = {2b\over X_0}\left( {v\over v_\perp }+{{v^2 }\over{v^2b +{\it GM} v_\perp}}v_c t\right).
\label{eq_length}
\end{equation}
The gap grows linearly with time from its post-encounter initial size, which in reality takes about an orbital period to develop. The gap that the extended mass profile of a sub-halo induces will have a similar character although the detailed functional form of the gap will depend on the mass profile of the perturber.

\subsubsection{Changes of Epicyclic Motions}

We  are also interested in the changes in the sizes of the epicycles. 
For an epicycle of radial extent $a$, $x(t) = a \cos{(\kappa t +\psi)}$, 
and the velocity $\dot{x}(t) =-\kappa a \sin{(\kappa t+\psi)}$ 
where $a$ and ${\psi}$ are the amplitude and orbital phase, respectively, of the epicycle. In the direction of the mean motion the epicyclic 
velocity is $\dot{y}(t) = -2\Omega a\cos{(\kappa t +\psi)}$ \citep{BT:08}. 
The epicycle is an ellipse, with axes in the ratio $2\Omega/\kappa$. 

An encounter with an initially completely cold stream will induce 
a systematic epicyclic motion through the combined effects of the velocity changes along and perpendicular to the stream.
The radial motion immediately after the encounter is $\Delta v_\perp(y)$. 
Along the stream there is a velocity change of $\Delta v_\parallel(y)$ to which
 we need to add the effective velocity change which results from the change to the new guiding center, Equation~\ref{eq_r}.
The change of mean guiding center leads to an angular velocity difference between the initial orbits and the post-encounter orbit which is simply the initial angular velocity,  
$v_c/X_0$, minus the new angular rotation rate, $\Omega(y)$, of Equation~\ref{eq_Om}. 
Expanding  Equation~\ref{eq_Om} to first order and multiplying by the
radius to get a linear velocity we find a second velocity change factor of  $\Delta v_\parallel(y)$. 
Therefore, to first order the new velocity along the stream, in the frame of the new guiding center
 is $2\Delta v_\parallel(y)$. 
Since $v_x/\kappa=a \sin{(\kappa t+\psi)}$ and $v_y/(2\Omega) = a \cos{(\kappa t+\psi)}$,
 the resulting added epicycle has a size in terms of the stream perturbations of,
\begin{equation}
a(y) = \sqrt{{{\Delta v^2_\perp(y)}\over \kappa^2} + {{\Delta v^2_\parallel(y)}\over \Omega^2}}.
\label{eq_a}
\end{equation}
The initial epicyclic phase, $\psi$, as a function of the stream position of this induced epicycle can be derived from the ratio of the two velocities at $t=0$,
\begin{equation}
\tan{(\psi(y))} = {2\Omega \Delta v_\perp(y) \over {\kappa \Delta v_\parallel(y)}}. 
\label{eq_psi}
\end{equation}
In a warm stream the pre-encounter epicycles will have random phases which the sub-halo induced epicycle will add leading to a spread in change in epicyclic size. Because the added epicyclic phase is coherent along the stream the stars will oscillate in and out together, with the stream varying from nearly straight, to a sideways reversed``S" shape. 

\section{Comparison to Idealized Simulations}

\subsection{Simulation Setup}

 Running some simple gravitational simulations of a star stream orbiting in a galactic potential that encounters a sub-halo with an extended mass profile tests the accuracy and the generality of the analytic predictions.
The galactic potential is an NFW halo \citep{NFW} with the peak of the rotation curve at 30 kpc 
where the circular velocity is 210 \kms that is a rough match to the dark halo of the Milky Way \citep{Aquarius} although those details are
not important for this illustrative calculation.
The radius and mass define a characteristic 
mass of $3.077 \times 10^{11} \msun$, which together give a set of characteristic scales
to normalize the numerical calculation units. 

We place 100,000 particles on an arc of length 1 radian on a circular orbit at 30 kpc with velocities equal to the local circular velocity. 
The arc is positioned so that a perturbing sub-halo will cross the middle of the stream.
Restricting attention to circular orbits is not a significant limitation for the relatively small size of the
gaps that sub-halos induce, although the gap size and density will expand and contract to conserve angular momentum
 if the stream is on a non-circular orbit.
One version of the  stream is completely cold with all particles having the same angular momentum, no radial velocities and  no initial width about the stream centerline. 
To understand the extension to warm streams with a finite width we also set up streams with
 a Gaussian distribution of epicycles about the same initial guiding centers. All the particles have  the same angular momentum to first order, so there is no shear in the stream. This allows direct comparison with the model predictions above, but will need to be taken into account for comparison to real streams.
In the warm simulations the epicycles are sufficiently small, generally less than 1\% of the orbital radius, 
that the asymmetric drift correction to the mean orbital velocity, which is a second order effect, is ignored. 

A sub-halo with a Hernquist mass profile having a mass of $10^6 \msun$, unless otherwise specified,
is sent on a straight line orbit towards the stream at a constant velocity of $1.2v_c$.   
The simulation is initiated with the sub-halo at a distance of one stream radius away from where 
it will cross the stream at $y=0$ at time zero.  The predictions for $\Delta v_\parallel(y)$ 
and $\Delta v_\perp(y)$ are worked out with the integrals of Equation~\ref{eq_dv}.
The results are only very weakly dependent on the precise angle with respect to the orbital plane
with which the sub-halo encounters the stream, but there are differences of detail.

\begin{figure}
\begin{center}
\includegraphics[angle=-90, scale=0.7]{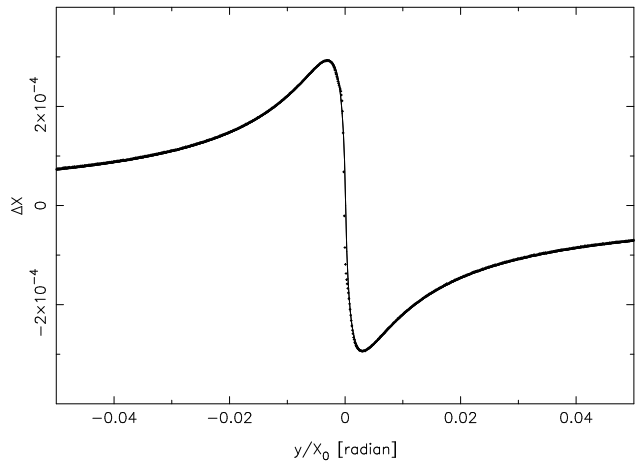}
\end{center}
\caption{The changes in the guiding center radius, normalized to 30 kpc,  after an encounter with a $10^6\msun$ sub-halo on a cold stream as a function of the pre-encounter co-ordinate along the stream. 
The impact parameter is 3 pc  whereas the sub-halo has a scale radius of about 100 pc. The line shows the predicted relation, which is nearly indistinguishable from the simulation points.}
\label{fig_rg0}
\end{figure}

\begin{figure}
\begin{center}
\includegraphics[angle=-90, scale=0.7]{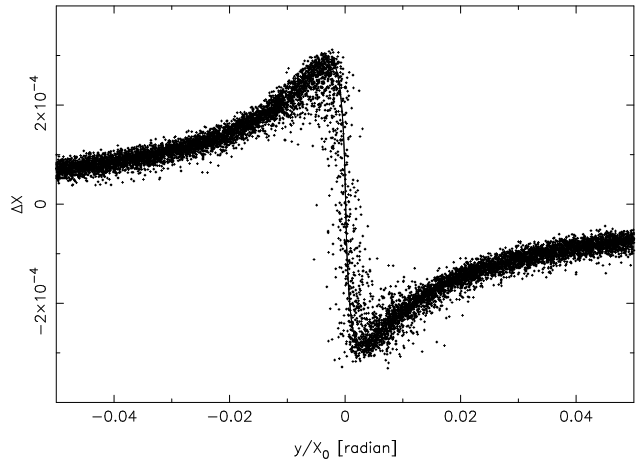}
\end{center}
\caption{Same as Figure~\ref{fig_rg0}, but for a stream of width 0.23\% of the orbital radius, equivalent to 70 pc.}
\label{fig_rg225}
\end{figure}

\begin{figure}
\begin{center}
\includegraphics[angle=-90, scale=0.7]{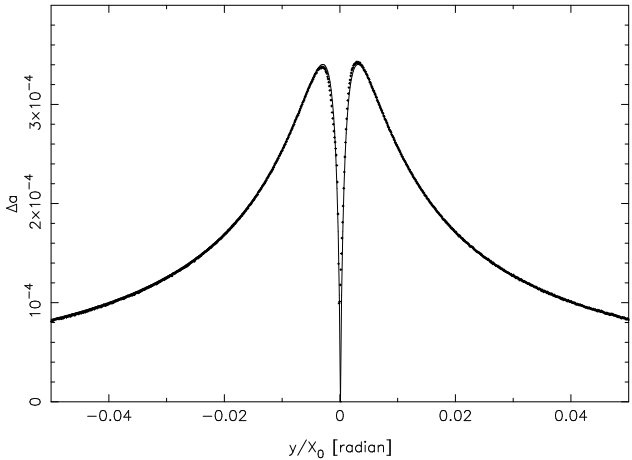}
\end{center}
\caption{The changes in the calculated epicycle size, normalized to 30 kpc, for the cold stream
simulation of Figure~\ref{fig_rg0}. 
The line shows the predicted relation which is nearly indistinguishable from the simulation points. 
In this case the sub-halo is shot from the outside in. There is a 
 slight asymmetry of the two peaks in the simulation results.}
\label{fig_e0}
\end{figure}

\begin{figure}
\begin{center}
\includegraphics[angle=-90, scale=0.7]{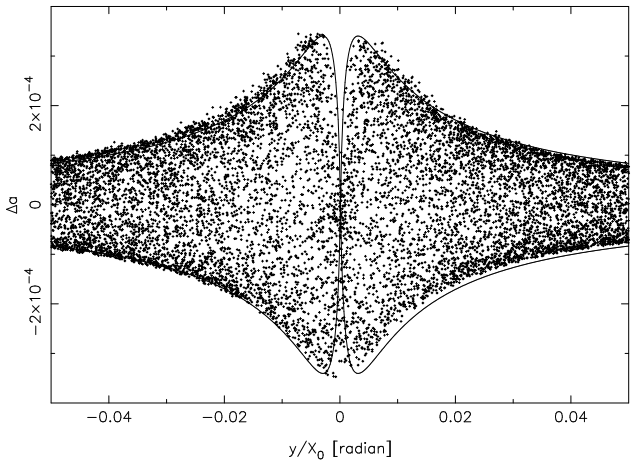}
\end{center}
\caption{The change in epicycle sizes for a warm stream of  width 70 pc. In this case the prediction  of Equation~\ref{eq_a} is shown as $\pm a(y)$. }
\label{fig_e225}
\end{figure}

We consider the response of completely cold and warm streams on circular orbits
 to a sub-halo of mass $10^6 \msun$ with a Hernquist mass profile
having a scale radius of 110 pc. 
For most encounters the impact parameter is set to 0.0001 of the orbital radius, that is,  3 pc, 
which is effectively a direct hit on the stream. The mass of $10^6 \msun$ is chosen on the basis of being a sufficiently small mass that the linear approximations in the analysis should in principle give good descriptions of the dynamics.  A significantly smaller sub-halo mass would be have a size scale smaller than the narrowest streams and of less practical interest since the gap may be too shallow to be readily visible. The steep LCDM mass spectrum of sub-halos, $N(M) \propto M^{-1.9}$ means that there be should be large numbers at these masses ensuring that real streams will have many encounters with masses in this range.  
 
The first simulated stream is completely cold and well matched to the 
assumptions of our analysis. 
The stream is warm with a width of  70 pc,  the width of the GD-1 star stream \citep{CG:13}, 
or 0.23\% of the orbital radius and is still very thin. A yet wider stream will be used to see where the approximations of our
analysis begin to lead to significant quantitative errors.

\subsection{Position and Velocity Changes in the Stream}

Figure~\ref{fig_rg0} shows the change in guiding center, $\Delta X=X-X_0$ for the cold stream along with the predicted relation
of Equation~\ref{eq_r}, integrated for the Hernquist potential used for the sub-halo in the simulation.
The prediction is virtually indistinguishable from the simulation points.
Figure~\ref{fig_rg225} shows the same prediction against the post-encounter guiding centers calculated from the angular momentum values  in a 70 pc wide stream. In this case there is a spread of values around the prediction, 
which we attribute to the fact that in a warm stream the individual stars have a range of velocities relative
to the sub-halo, which leads to a range of relative $v_\parallel$ and $v_\perp$ for individual particles. 
For yet warmer streams (not shown) there is even more scatter and the
angular momentum changes can even reverse sign, as anticipated  
in our analytic results for the point mass object, Equation~\ref{eq_ratio}.

The change in the sizes of the epicycles, $\Delta a=a-a_0$, where $a$ and $a_0$ are the post encounter and initial epicyclic sizes, respectively,  are compared to the prediction of Equation~\ref{eq_a} for the cold stream 
(where $a_0=0$) in Figure~\ref{fig_e0} showing that
the prediction is quite good.
The height of the two peaks in the $\Delta a$ distribution in the simulation are slightly  asymmetric
 whereas the prediction is completely symmetric about the center point. To identify the source of the asymmetry some variants on the
basic simulation were done. The plot of Figure~\ref{fig_e0} uses a sub-halo started at twice the orbital radius and was shot inward. 
We repeated the simulation with the sub-halo shot outward from the center, finding that the asymmetry reverses and become stronger. 
The time evolution of $\Delta a$ shows that before the sub-halo arrives at the stream $\Delta a$ rises with time approximately linearly
with distance along the stream, being strongest in the leading part of the stream, to which the sub-halo was closest as it approached the crossing point. 
Starting the sub-halo closer to the stream reduces the size of the pre-crossing $\Delta a$ and  the final asymmetry. Larger mass sub-halos do not lead to proportionally larger asymmetries. 
We conclude that the approximation of a straight line stream for the calculation of the velocity perturbations, which are
 then translated into the motions of a stream on a circular orbit does not capture all of the dynamics, but the approximation remains good in this case to a few percent.

 For the warm stream the changes in epicycle sizes, $\Delta a$,  are displayed in Figure~\ref{fig_e225}.  
The perturbation
induced epicycles are adding to pre-existing epicycles with random phases, 
hence there will be a spread from completely in-phase addition to out of phase subtraction. Figure~\ref{fig_e225} shows that the
predicted relation and its negative largely bound the outcomes. 

\begin{figure}
\begin{center}
\includegraphics[angle=-90, scale=0.8]{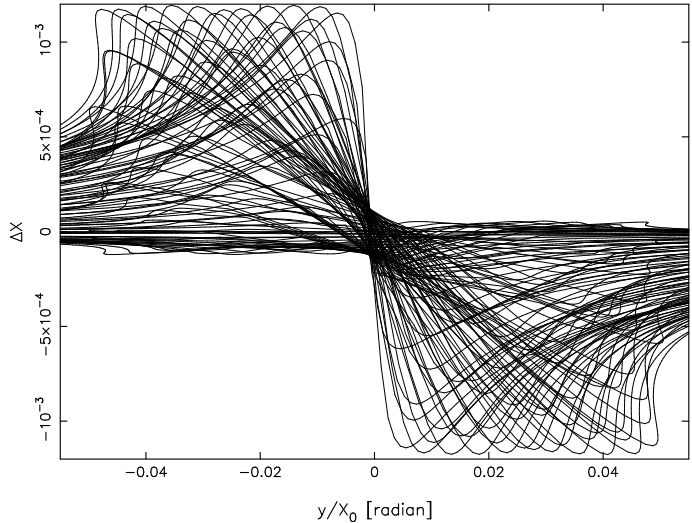}
\end{center}
\caption{The evolution with time of the $xy$ location of particles in a cold stream after a $3\times 10^6 \msun$ object passes through. 
The initial time has the shape of Figure~\ref{fig_rg0} and the late times become saw-tooth shaped.
Note that the $x$ scale is greatly magnified relative to the $y$ scale. The stream is shown every 0.8 units of time (about 1/8 of an orbital period) for 100 times. Line segments are drawn between a subset of the simulation points.}
\label{fig_xy}
\end{figure}

The evolution in time of the $xy$ location  of the particles in a cold stream simulation, 
effectively equivalent to the guiding centers of a warm stream, is shown in Figure~\ref{fig_xy}. 
For this example we use a sub-halo of mass $3\times 10^6$ which increases the change in
shape of the perturbed region with time.
The stream starts with the shape of Figure~\ref{fig_rg0}, however with time the outer part falls behind the inner part.  
The characteristic sideways reversed ``S" shape gradually becomes closer to a saw-tooth shape. 
Such shapes may be marginally visible in some current stream data \citep{CG:13} and may become useful sub-halo detection signals in themselves as the quality of stream data improves to allow surface density measurements, not just linear densities along the stream.

\subsection{Gaps in Streams}

\begin{figure}
\begin{center}
\includegraphics[angle=-90, scale=0.7]{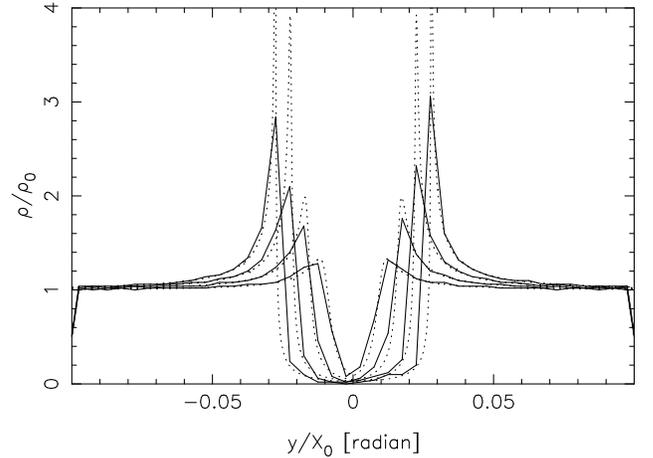}
\end{center}
\caption{The line show the development of a density gaps in a nearly completely cold stream vs the angular coordinate after an encounter with a $10^6\msun$ sub-halo at dimensionless times 20, 40, 60 and 80, where the rotation period is $2\pi$.
The dotted lines show the predicted relation at different times.}
\label{fig_d1}
\end{figure}

The development of the gap in the density of  a stream according to Equations~\ref{eq_den} and \ref{eq_phit}
is compared to a cold stream stimulation in Figure~\ref{fig_d1}. The predictions does a very good job of describing the shape 
of the gap and its growth with time.

\begin{figure}
\begin{center}
\includegraphics[angle=-90, scale=0.7]{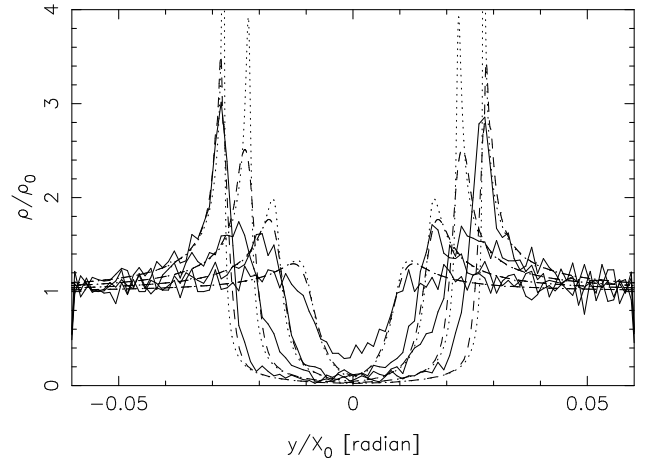}
\end{center}
\caption{Same as Figure~\ref{fig_d1} except for a warm stream of width 70 pc. The dashed line shows the effect of adding in 
the epicyclic smoothing of the stream width along the stream.}
\label{fig_d00225}
\end{figure}

Warm streams have epicyclic motions which blur the density out along the stream which is not yet
included in our cold stream analysis.
Figure~\ref{fig_d00225} shows the outcome of the same small impact parameter encounter as in
Figure~\ref{fig_d1} but now on a warm stream of width 70 pc. 
The cold stream prediction (dotted line) does not do a very good job. 
However, we modify the prediction by simply convolving the predicted density distribution with a Gaussian.
The spread $\sigma$ of the Gaussian is related to the FWHM of the stream multiplied by $1.414/2.355$ where the factor of 1.414 allows for the fact that the epicycles  along the stream are larger than their perpendicular size in the ratio $2\Omega/\kappa$, which is $\sqrt{2}$ for a flat rotation curve, and, the factor $2.355$ converts from FWHM to a Gaussian width.

\begin{figure}
\begin{center}
\includegraphics[angle=-90, scale=0.7]{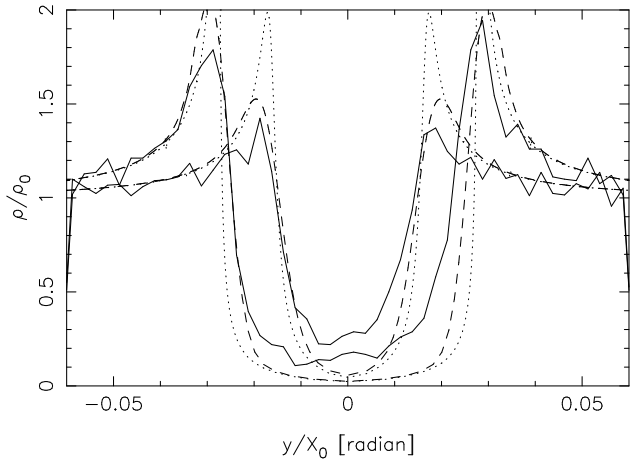}
\end{center}
\caption{Same as Figure~\ref{fig_d1} except for a warm stream of width 150 pc. The dashed line shows the effect of adding in 
the epicyclic smoothing of the stream width along the stream. For clarity only the times 40 and 80 are plotted and the vertical scale has been reduced.}
\label{fig_d005}
\end{figure}

We compare the gap measurements to the predictions for a 70 pc wide stream in Figure~\ref{fig_d00225}.
 The cold stream prediction (dotted line) gives a deeper and sharper gap than is found in the simulation (solid line). 
The velocity dispersion allowance does quite a good job predicting the shape of the gap (dashed line).  
Figure~\ref{fig_d005} shows a stream of width 150 pc where the width corrected prediction of the overall shape remains qualitatively correct, but a significant asymmetry is developing and the predicted depth is greater than measured. 
Not shown is a 300 pc wide stream, 1\% of the orbital radius, which continues the trend of increasing discrepancy
from our predictions with increasing stream width. 
The most significant error is that the depth is measured to be 0.5 to 0.6 of the mean level, whereas the predictions go nearly to 100\% depth.
The prediction continues to provide a useful estimate of gap width, but does not give an accurate density profile.

\subsection{The Effect of Shear in Streams}

The individual stars in stellar streams are drawn out from their progenitor objects with a range of angular momenta and radial velocities \citep{YJH:11, EB:11} which together fix the width of the stream. 
Both the angular momentum and the radial action, $J_r=\kappa a^2$ are conserved quantities in an axisymmetric potential, which 
applies once the stars are away from the progenitor object. 
Progenitors that lose stars through a fairly well collimated outflow through the Lagrange points will have stream that are
largely $J_r$ dominated, with relative little shear \citep{Kupper:08, Kupper:12, EB:11}.

If there is significant range of angular momentum in the stream then it will cause gaps to shear with time so that any one-dimensional measure will see a decreased gap density once the tilt of the gap exceeds the gap width. 
The spreading of the gap into a tilted stripe along the stream and the consequent reduction of the depth at any 
location is straightforward to calculate. 
However, a useful prediction needs to have some way to estimate the range of angular momentum that is likely to be present. 
Such a prediction requires full n-body simulations that take into account the details of how the progenitor dissolves as it orbits in the potential of its host galaxy, which is beyond the scope of this paper. 

A simple calculation shows when shear becomes an important effect. 
For illustration we assume that in a 60 pc wide stream 
half of the width at the origin of the stream is due to angular momentum \citep{EB:11}.
In this case  the range of angular momentum relative to the mean is 0.001 at 30 kpc orbital radius. 
Since small gaps are blurred out by the velocity dispersion in the stream,  we 
consider the evolution of a gap having a length twice the width of the stream, 120 pc, 0.004 times the orbital radius.
The stream shear in this case  will reduce the density contrast in this gap to about half its initial depth in 
the time it takes the low and high end of the angular momentum in the stream to rotate apart by about
twice the initial width of the gap. 
That time is 2 times the initial gap width divided by the fractional shear, or, 
$2\times 0.004/0.001$ which is 8 units of time, or about 1.3 rotation periods, which at 30 kpc is about 1 Gyr.
The range of angular momentum spread diminishes linearly with distance down the stream, so at half the stream length the gap would take 2 Gyr to diminish to half its depth. 
Wider gaps would blur out over times directly proportionally to their length, so
a gap ten times the stream width half way down the stream would last 10 Gyr.
The simulations of dissolving globular clusters by \citet{Kupper:08, Kupper:12} suggest 
that the angular momentum range in those cases may be so small that angular momentum smearing is not a significant effect.
We emphasize that the range of angular momentum relative to the radial action in the stars lost through the tidal lobes 
has not been extensively studied and the question of how both shear and substantially elliptical stream orbits 
effect the visibility of gaps remains to be resolved.

\section{DISCUSSION and CONCLUSIONS}

This paper provides a first order dynamical analysis of the velocity changes and subsequent development of a gap in a star stream
after a massive object passes. A gap with a length comparable to the stream width
 will be blurred out through the subsequent orbital motions within the stream. Gaps that survive 
need to be 2 to 5 times the stream width, depending on details of the orbit and how stars are unbound from the progenitor
object.
The theory provides useful quantitative predictions for the resulting development 
of gaps in a stream, provided that the stream width is less than about 1\% of its mean orbital radius.
The numerical model can be used to make predictions of gap shapes in a variety of orbital situations.
 One insight is that although the single shape approach to gap-filtering developed in \citet{CGH:12,CG:13} 
 is a reasonable approximation, the density profile gaps vary widely in their sharpness of shoulders. 

The limitation of this study is that we study circular stream orbits and 
the details of the angular momentum, radial action and phase angle
distributions of stars once unbound from the progenitor system is not yet sufficiently well understood in
general situations to be incorporated into the model.
Those details have an important influence on the real space appearance of streams and are the focus of ongoing studies.
 
\acknowledgements

This research was supported by CIFAR and NSERC Canada. An anonymous referee and Wayne Ngan provided comments that have
strengthened and clarified the paper.


\begin{thebibliography}{99}


\bibitem[Angulo et al.(2013)]{AHA:13} Angulo, R.~E., Hahn, O., \& Abel, T.\ 2013, arXiv:1304.2406 

\bibitem[Benson et al.(2013)]{Benson:13} Benson, A.~J., Farahi, A., Cole, S., et al.\ 2013, \mnras, 428, 1774 

\bibitem[Binney \& Tremaine(2008)]{BT:08} Binney, J., \& Tremaine, S.\ 2008, Galactic Dynamics: Second Edition, Princeton University Press

\bibitem[Barkana et al.(2001)]{BHO:01} Barkana, R., Haiman, Z., \& Ostriker, J.~P.\ 2001, \apj, 558, 482 

\bibitem[Bode et al.(2001)]{Bode:01} Bode, P., Ostriker, J.~P., \& Turok, N.\ 2001, \apj, 556, 93 
\bibitem[Carlberg(2009)]{Carlberg:09} Carlberg, R.~G.\ 2009, \apjl, 705, L223 
\bibitem[Carlberg(2012)]{Carlberg:12} Carlberg, R.~G.\ 2012, \apj, 748, 20 
\bibitem[Carlberg et al.(2012)]{CGH:12} Carlberg, R.~G., Grillmair, C.~J., \& Hetherington, N.\ 2012, \apj, 760, 75 
\bibitem[Carlberg \& Grillmair(2013)]{CG:13} Carlberg, R.~G. \&  Grillmair, C.~J.\ 2013,  \apj, 768, 171 
\bibitem[Dehnen et al.(2004)]{DOGR:04} Dehnen, W., Odenkirchen, M., Grebel, E.~K., \& Rix, H.-W.\ 2004, \aj, 127, 2753 
\bibitem[Diemand, Kuhlen \& Madau(2007)]{VL1} Diemand J., Kuhlen M., Madau P., 2007, \apj, 667, 859
\bibitem[Eyre \& Binney(2011)]{EB:11} Eyre, A., \& Binney, J.\ 2011, \mnras, 413, 1852 

\bibitem[Johnston(1998)]{Johnston:98} Johnston, K. V. 1998, \apj, 495, 297


\bibitem[Johnston et al.(2001)]{JSB:01} Johnston, K.~V., Sackett, P.~D., \& Bullock, J.~S.\ 2001, \apj, 557, 137 

\bibitem[Johnston, Spergel \& Haydn(2002)]{JSH:02} Johnston, K. V., Spergel, D. N. \& Haydn, C. 2002, \apj, 570, 656

\bibitem[Ibata et al.(2002)]{Ibata_etal:02} Ibata, R.~A., Lewis, G.~F., Irwin, M.~J., \& Quinn, T.\ 2002, \mnras, 332, 915 

\bibitem[K{\"u}pper et al.(2008)]{Kupper:08} K{\"u}pper, A.~H.~W., MacLeod, A., \& Heggie, D.~C.\ 2008, \mnras, 387, 1248 
\bibitem[K{\"u}pper et al.(2010)]{Kupper:10} K{\"u}pper, A.~H.~W., Kroupa, P., Baumgardt, H., \& Heggie, D.~C.\ 2010, \mnras, 401, 105 
\bibitem[K{\"u}pper et al.(2012)]{Kupper:12} K{\"u}pper, A.~H.~W., Lane, R.~R., \& Heggie, D.~C.\ 2012, \mnras, 420, 2700 

\bibitem[Navarro et al.(1997)]{NFW} Navarro, J.~F., Frenk, C.~S., \& White, S.~D.~M.\ 1997, \apj, 490, 493 

\bibitem[Schneider et al.(2013)]{SSR:13} Schneider, A., Smith, R.~E., \& Reed, D.\ 2013, \mnras, 1489 

\bibitem[Siegal-Gaskins \& Valluri(2008)]{SGV:08} Siegal-Gaskins, J.~M., \& Valluri, M.\ 2008, \apj, 681, 40 
\bibitem[Springel et al.(2008)]{Aquarius}Springel, V. et al. 2008, \mnras, 391, 1685 
\bibitem[Stadel et al.(2009)]{Stadel:09} Stadel, J., Potter, D., Moore, B., et al.\ 2009, \mnras, 398, L21 
\bibitem[Yoon, Johnston \& Hogg(2011)]{YJH:11} Yoon, J. H., Johnston, K. V., \& Hogg, D. W. 2010, \apj, 731, 58

\end{thebibliography}
\end{document}